\documentclass[10pt]{article}
\usepackage{graphicx}
\usepackage{layout}

\begin{document}

\author{A. Tartaglia, M.L. Ruggiero \\
Dip. Fisica, Politecnico, and INFN Torino, Italy}
\title{Gravitomagnetic measurement of the angular momentum of celestial
bodies}
\maketitle

\begin{abstract}
The asymmetry in the time delay for light rays propagating on opposite sides
of a spinning body is analyzed. A frequency shift in the perceived signals
is found. A practical procedure is proposed for evidencing the asymmetry,
allowing for a measurement of the specific angular momentum of the rotating
mass. Orders of magnitude are considered and discussed.
\end{abstract}

\section{Introduction}

A well known effect of gravity on the propagation of electromagnetic signals
is the time delay: for example a radar beam emitted from a source on Earth
toward another planet of the solar system, and hence reflected back,
undergoes a time delay during its trip (with respect to the propagation in
flat space-time), due to the influence of the gravitational field of the
Sun. This effect was indeed called the "fourth" test of General Relativity%
\cite{shapiro64},\cite{shapiro66}, coming fourth after the three classical
ones predicted by Einstein\cite{einstein}.

Though small, the time delay in the propagation of electromagnetic waves was
detected by Shapiro et al. \cite{shapiro}, timing radar echoes from Mercury
and Venus, by means of the radio-telescopes of Arecibo and Haystack.
Anderson et al. \cite{ander} measured the time delay of the signals
transmitted by Mariner 6 and 7 orbiting around the Sun. Finally Shapiro and
Reasenberg obtained more accurate results using a Viking mission that
deposited a transponder on the surface of Mars: the theoretical prediction
was verified within $\pm 0.1\%$ \cite{marte1} \cite{marte2}.

These measurements accounted just for the presence of a massive source,
described by the Schwartzschild solution of the Einstein field equations. In
this paper we are going to work out another correction to the time delay,
due to the spin of the source, which is produced by the gravitomagnetic
interaction. In fact, the off-diagonal term of the metric tensor around a
rotating body, produces a correction to the time delay, which has opposite
signs on opposite sides of the body. The asymmetry, within the solar system,
is in any case small, as we shall see: however its systematicness lends an
opportunity to reveal the effect, appropriately combining the ticks of a
'clock' passing behind the spinning mass. In fact the relative motion of
source, receiver and central mass, produces a varying time delay, which
shows up as a small frequency shift; the gravitomagnetic contribution to
this shift is manifested as an asymmetry between right and left with respect
to the central mass. In these conditions, if one superposes corresponding
records, before and after the occultation of the clock, a beating function
will result, where the magnitude of the frequency of the beats is
proportional to the angular momentum of the spinning body, and the frequency
of the basic signal is shifted with respect to the flat space-time situation
by an amount proportional to the mass.

Considering the actual orders of magnitude inside the solar system, we shall
show that the effect is not far from the threshold of detectability.

\section{The time delay}

Let us review the propagation of electromagnetic signals in a stationary
axially symmetric space time, i.e. in the vicinity of a massive rotating
body. If we confine our analysis to the 'equatorial plane' of the system a
null world line corresponds to the equation

\begin{equation}
0=g_{tt}dt^{2}+g_{rr}dr^{2}+g_{\phi \phi }d\phi ^{2}+2g_{t\phi }dtd\phi
\label{null}
\end{equation}
In a flat space time the coordinates would be the usual polar ones. The
symmetry tells us that the $g$ functions depend on $r$ and $\theta $ only,
which in practice in our case means on $r$ only ($\theta =\pi /2$).

From (\ref{null}) one has

\[
dt=\frac{-g_{t\phi }\pm \sqrt{g_{t\phi }^{2}-g_{tt}\left( g_{rr}\left( \frac{%
dr}{d\phi }\right) ^{2}+g_{\phi \phi }\right) }}{g_{tt}}d\phi
\]

If we use the axis of the angular momentum of the central mass as a
reference for the positive rotation direction, we have to choose the $+$
sign when $d\phi >0$ and the $-$ sign when $d\phi <0$. Let us say the first
condition corresponds to be on the left of the mass, the second one to be on
the right. Then
\begin{eqnarray}
dt_{l} &=&\frac{-g_{t\phi }+\sqrt{g_{t\phi }^{2}-g_{tt}\left( g_{rr}\left(
\frac{dr}{d\phi }\right) ^{2}+g_{\phi \phi }\right) }}{g_{tt}}d\phi
\label{diffe} \\
dt_{r} &=&\frac{g_{t\phi }+\sqrt{g_{t\phi }^{2}-g_{tt}\left( g_{rr}\left(
\frac{dr}{d\phi }\right) ^{2}+g_{\phi \phi }\right) }}{g_{tt}}\left| d\phi
\right|  \nonumber
\end{eqnarray}

Integrating (\ref{diffe}) along geodesic arcs (expressed as functions $%
r\left( \phi \right) $) one obtains the corresponding coordinated times of
flight. In a curved space time the result will in general be greater than in
a flat one.

Remarkably we see also that, for the same geometric path on the left and on
the right of the central mass, the results will be different because of
gravitomagnetic effects (the ones induced by the off-diagonal term of the
metric tensor). The difference in time differentials for equal azimuthal
span and trajectory on the two sides is
\begin{equation}
dt_{lr}=2\frac{g_{t\phi }}{g_{tt}}\left\vert d\phi \right\vert
\label{asimme}
\end{equation}
Integrating (\ref{asimme}) along a null geodesic line over a finite range of
$\phi $ values gives a time of flight asymmetry between right and left. This
asymmetry could have some relevance in gravitational lensing phenomena \cite%
{ciufo}.

\section{Weak field condition}

\begin{figure}[tp]
\begin{center}
\includegraphics[width=11cm,height=10cm]{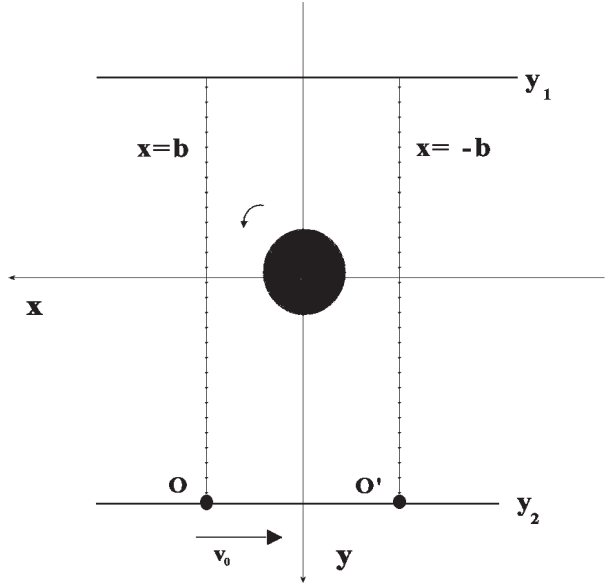}
\end{center}
\caption{{\protect\small Configuration of the problem. The trajectory of the
signals is assumed to be straight. A Cartesian coordinate system, centered
on the spinning mass, is used. The source is at }$y_{1}$, the observer $O$
is at $y_{2}$. The observer is moving with the velocity $v_{0}$. Two
symmetric situations (at different times) are considered.}
\label{fig:figura}
\end{figure}

We now specialize our analysis to a weak field condition such as the one we
find within the solar system. In this case Cartesian (i.e. rectangular)
coordinates are often used. Assuming that the $z$ axis coincides with the
direction of the angular momentum three-vector (\ref{null}) becomes
\begin{equation}
ds^{2}=g_{tt}dt^{2}+g_{xx}dx^{2}+g_{yy}dy^{2}+g_{zz}dz^{2}+2g_{xt}dxdt+2g_{yt}dydt
\label{eq:metrica1}
\end{equation}
The $g$'s now depend on $x$ and $y$.

The weak field hypothesis is commonly used to neglect the effect on the time
of flight, of the bending of the trajectory due to the central mass \cite%
{straumann}. This assumption, appropriately choosing the $x$ and $y$ axes,
leads to a ray trajectory that is a straight line $x=b=const$; $b$ is of
course the closest approach distance with respect to the spinning body. We
must add the explicit form of the weak field metric elements
\begin{eqnarray*}
g_{yt} &=&G\frac{2Max}{c\left( x^{2}+y^{2}\right) ^{\frac{3}{2}}} \\
g_{tt} &=&c^{2}-G\frac{2M}{\sqrt{x^{2}+y^{2}}} \\
g_{yy} &=&-1-G\frac{2M}{c^{2}\sqrt{x^{2}+y^{2}}}
\end{eqnarray*}
Here $M$ is the mass of the central body, $J$ is its angular momentum, and $%
a=\frac{J}{Mc}$.

Under these conditions the time of flight of the electromagnetic signals
becomes
\begin{equation}
t_{f}(y_{1},y_{2})=t_{0}+t_{M}+t_{J}  \label{eq:delayc}
\end{equation}
where (see \cite{ruffi} and \cite{ruggierotartaglia02})
\begin{eqnarray}
t_{0} &=&\frac{y_{2}-y_{1}}{c}  \label{eq:tzero} \\
t_{M} &=&\frac{2GM}{c^{3}}\ln \frac{y_{2}+\sqrt{b^{2}+y_{2}^{2}}}{y_{1}+%
\sqrt{b^{2}+y_{1}^{2}}}  \label{eq:temme} \\
t_{J} &=&\mp \frac{2GMa}{c^{3}b}\left[ \frac{y_{2}}{\sqrt{b^{2}+y_{2}^{2}}}-%
\frac{y_{1}}{\sqrt{b^{2}+y_{1}^{2}}}\right]  \label{eq:temmea}
\end{eqnarray}
The quantity $y_{1}$ is the $y$ coordinate of the source of the signals, $%
y_{2}$ is the $y$ coordinate of the receiver. The time $t_{0}$ is clearly
the Newtonian time of flight; $t_{M}$ is the known gravitational time delay
(Shapiro time delay), and $t_{J}$ is the correction to the time delay
produced by the gravitomagnetic interaction with the angular momentum of the
central body.\newline
The double sign in $t_{J}$ means that gravitomagnetism shortens the time of
flight on the left and lengthens it on the right.

\section{Variable impact parameter}

In general $y_{1}$ and $y_{2}$ will not be fixed and, consequently, the
whole time delay will vary in time. Let us simplify the situation assuming
that the source of the electromagnetic radiation is much more distant from
the spinning body than the receiver. The source is then pointing out a fixed
direction in space, just as a far astronomical source would do, and the time
variation of $y_{2}$ in the reference frame of the central mass may easily
be converted in a time variation of $b$ from the viewpoint of the observer
(receiver).

If we limit the consideration to a situation around the occultation of the
source by the central body, the source and the observer will be almost
opposed with respect to the center, and the time dependence of $b$ on $t$
will be approximately linear. From an arbitrary (but not too big with
respect to the radius of the central body) distance $b_{0}$, and considering
the approach to the occultation, we can write\footnote{%
At least for short enough time, the linear approximation of eq. (\ref{bi})
is valid. As we say below, in a more realistic situation, the relative
motions of source, rotating body and observer must be taken into account.}
\begin{equation}
b\simeq b_{0}-v_{0}t  \label{bi}
\end{equation}%
where $v_{0}$ is a positive velocity (apparent transverse velocity of the
source in the sky).

If we add the condition that the time scale of the change of $b$ with time
is much bigger than the period $T$ of the incoming electromagnetic wave, we
can expect the period to slowly change, in one period time, by the amount
\begin{equation}
\delta T=\frac{dt_{f}}{db}\frac{db}{dt}T  \label{deltat}
\end{equation}
Explicitly performing the differentiations and letting $y_{1}\rightarrow
-\infty $%
\begin{eqnarray}
\frac{\delta T}{T} &=&-\frac{\delta \nu }{\nu }=\frac{2GM}{c^{2}}\frac{%
b^{2}+2\sqrt{b^{2}+y_{2}^{2}}y_{2}+2y_{2}^{2}}{b\sqrt{b^{2}+y_{2}^{2}}\left(
y_{2}+\sqrt{b^{2}+y_{2}^{2}}\right) }\frac{v_{0}}{c}  \nonumber \\
&&\mp 2\left( y_{2}^{3}+2b^{2}y_{2}+\left( b^{2}+y_{2}^{2}\right) ^{\frac{3}{%
2}}\right) GM\frac{a}{c^{2}b^{2}\left( b^{2}+y_{2}^{2}\right) ^{\frac{3}{2}}}%
\frac{v_{0}}{c}  \label{deltrela}
\end{eqnarray}

Now $\nu $ represents the frequency of the signals.

A further reasonable assumption is that $b<<y_{2}\sim R$, where $R$ is the
distance of the observer from the central body. Under this assumption and
posing $\mu =G\frac{M}{c^{2}}$ (\ref{deltrela}) can be simplified to
\begin{equation}
\frac{\delta \nu }{\nu }=\allowbreak 4\frac{\mu }{b}\frac{v_{0}}{c}-\frac{%
\mu b}{R^{2}}\allowbreak \frac{v_{0}}{c}\allowbreak \mp 4\frac{\mu a}{b^{2}}%
\frac{v_{0}}{c}  \label{red}
\end{equation}

As it can be seen, there is an effective frequency shift, which is red
during the approach. The last contribution has different signs on the
opposite sides, thus leading to an asymmetry in the frequency shift too.

Possible asymmetries caused by the gravitomagnetic influence of the Sun on
the propagation of electromagnetic signals were indeed studied by Davies
\cite{davies}, who proposed an experiment using a pair of sattelites
orbiting the Sun. The idea was to measure the time of flight difference
between a clockwise and a counter-clockwise trajectory of electromagnetic
signals going from the Earth to one of the satellites, then to the other,
then back. Our proposal is different in that we are considering effects on
the frequency of the signals.

Another approach to the asymmetries due to the angular momentum of the Sun
was considered by Bertotti and Giampieri\cite{bertotti}. There a complete
analysis of the Doppler shift led to a result including the effect we have
worked out. Here we propose a specific method to obtain the relevant
observable.

\section{Evidencing the asymmetry}

Let us assume that the electromagnetic signal is a pure harmonic wave, whose
proper frequency is $\nu _{0}$. At a given time, not far from the
occultation, the received frequency will be $\nu _{0}+\delta \nu =\nu
_{0}\left( 1+\alpha _{1}\mp \alpha _{2}\right) $ where $\alpha _{1}$ is the
symmetric part of the shift, and $\alpha _{2}$ is the antisymmetric one.

Suppose now to be able to record the received signal for a given time span,
from the $b_{0}$ line of sight distance from the central body up to the
occultation, then again from the reappearance on the other side up to a
symmetric line of sight distance $b_{0}$. The second recorded signal can be
reversed in time, then superposed to the one corresponding to the approach
interval. In symbols all this amounts to generate a compounded beating
signal:
\begin{eqnarray}
\xi &=&\xi _{0}\left( \sin \left( 2\pi \nu _{0}\left( 1+\alpha _{1}-\alpha
_{2}\right) t\right) +\sin \left( 2\pi \nu _{0}\left( 1+\alpha _{1}+\alpha
_{2}\right) t\right) \right)  \label{batti} \\
&=&2\xi _{0}\cos \left( 2\pi \nu _{0}\alpha _{2}t\right) \sin \left( 2\pi
\nu _{0}\left( 1+\alpha _{1}\right) t\right)  \nonumber
\end{eqnarray}
Here $\xi _{0}$ is the constant amplitude of the original signal.

The resulting signal has a (slowly) time varying frequency
\begin{equation}
\nu _{1}=\nu _{0}\left( 1+4\frac{\mu }{b}\frac{v_{0}}{c}-\frac{\mu b}{R^{2}}%
\allowbreak \frac{v_{0}}{c}\right)  \label{portante}
\end{equation}
and a modulated amplitude, whose modulation frequency is (again with a slow
change in time)
\begin{equation}
\nu _{2}=4\nu _{0}\frac{\mu a}{b^{2}}\frac{v_{0}}{c}  \label{modula}
\end{equation}

\section{Numerical estimates}

To fix a few numbers, we can consider the situation in the solar system,
with the Sun as the central spinning body, and an Earth bound observer. The
source is a far away astronomical body (e.g. a pulsar). In this case the
orders of magnitude are
\[
\begin{array}{ccccc}
\mu \sim 10^{3}\mathit{\ m} & a\sim 10^{3}\mathit{\ m} & b\sim 10^{9}\mathit{%
\ m} & R\sim 10^{11}\mathit{\ m} & v_{0}\sim 10^{4}\mathit{\ m/s}%
\end{array}%
\]

As a consequence we have
\begin{eqnarray}
4\frac{\mu }{b}\frac{v_{0}}{c} &\sim &10^{-10}  \nonumber \\
\frac{\mu b}{R^{2}}\allowbreak \frac{v_{0}}{c} &\sim &10^{-14}  \label{stime}
\\
4\frac{\mu a}{b^{2}}\frac{v_{0}}{c} &\sim &10^{-16}  \nonumber
\end{eqnarray}

The order of magnitude of the effective red shift is extremely small.

As for the beats, the smallness of their frequency implies a long time for
significant changes in the amplitude of the resulting signal. The phase of
the modulating wave is
\[
\phi =8\pi \frac{\mu a}{b^{2}}\frac{v_{0}}{c}\nu _{0}t\sim 10^{-16}\nu _{0}t
\]
which means that a full cycle happens when $\nu _{0}t\sim 10^{16}$. This
condition requires a longer or shorter time according to the frequency of
the signal: for optical frequencies 0.1 s would be enough; in the GHz range $%
10^{7}$s would be needed. Consider for comparison that the time required to
move across the sky by one solar radius is, in these conditions, $\sim
10^{5} $ s; this means that a non-negligible effect on the amplitude could
be seen using a signal in the GHz range and tracking it for $\sim 10^{5}$ s
before and after the occultation of the source by the Sun.

\section{Conclusion}

We have evidenced a frequency effect connected with a varying time delay in
the propagation of electromagnetic signals in the vicinity of a spinning
massive body. The effect, if considered in solar system conditions is, as
usual, very small, however not entirely negligible, at least when the trick
proposed in the text is used, i.e. producing a beat between signals after
and before the occultation. Of course many practical problems have to be
considered and discussed to transform some principle formulae into an actual
measurement. First of all one has to find the way to record the signals
before and after an occultation event. Any recording device should contain a
clock at a frequency higher than the one to be recorded; this seems to place
a limit at a few GHz. Then the length of the record should not exceed the
coherence time of the source, and this depends on the nature of the source.
Furthermore there is also a problem with the amplitude and frequency
stability of the signal. If the source is an actual clock on board a
spacecraft, the formulae expressing the time delay are also geometrically
more complicated, because of the not negligible motion of the source.

We can also envisage more favorable conditions. In case, for instance, we
could find a pulsar orbiting a neutron star, the minimum impact parameter $b$
would be $\sim 10^{4}$ m. Keeping the same values for the rest, the
asymmetric contribution to the frequency shift (last term in \ref{red} and %
\ref{stime}) would be $\sim 10^{-6}$, which is indeed easy to detect. For a
typical frequency of the signal $10^{3}$ Hz the time needed to fully measure
the beats we have described in sect. 5 would be $\sim 10^{3}$ s (20 minutes).

The conclusion we can draw at this point is that the method we propose seems
promising for detecting gravitomagnetic effects on the propagation of
electromagnetic signals, and for measuring the angular momenta of
astronomical bodies.

\end{document}